\documentclass[oupdraft]{bio}
\usepackage{enumitem}
\usepackage{hyperref}
\usepackage{multirow}

\usepackage{amsmath}
\usepackage{amsfonts}
\usepackage{amssymb}
\usepackage{bm}
\usepackage{color}
\usepackage{ulem}

\usepackage{verbatim}

\usepackage{times}

\history{Received August 1, 2010;
revised October 1, 2010;
accepted for publication November 1, 2010}

\begin{document}

\title{Bayesian Inference of Networks Across Multiple Sample Groups and Data Types}

\author{ELIN SHADDOX$^\ast$, CHRISTINE B. PETERSON, FRANCESCO C. STINGO, NICOLA A. HANANIA, CHARMION CRUICKSHANK-QUINN, KATERINA KECHRIS, RUSSELL BOWLER, MARINA VANNUCCI$^\ast$\\[4pt]
\textit{
$^{1}$Dept of Statistics, Rice University, Houston, TX, USA\\
$^{2}$Dept of Biostatistics, UT MD Anderson Cancer Center, Houston, TX, USA\\
$^{3}$Dept of Statistics, Computer Science, Applications ``G. Parenti", University of Florence, Florence, Italy\\
$^{4}$Dept of Medicine-Pulmonary, Baylor College of Medicine, Houston, TX, USA\\
$^{5}$Dept of Pharmaceutical Sciences, School of Pharmacy, University of Colorado, Denver, CO, USA\\
$^{6}$Dept of Biostatistics \& Informatics, Colorado SPH, University of Colorado, Denver, CO, USA\\
$^{7}$Dept of Medicine, National Jewish Health, Denver, CO, USA}\\
{elin@rice.edu, marina@rice.edu}}

\markboth%
{E. Shaddox and others}
{Bayesian Inference of Networks}

\maketitle

\footnotetext{To whom correspondence should be addressed.}

\begin{abstract}
{In this paper, we develop a graphical modeling framework for the inference of networks across multiple sample groups and data types. In medical studies, this setting arises whenever a set of subjects, which may be heterogeneous due to differing disease stage or subtype, is profiled across multiple platforms, such as metabolomics, proteomics, or transcriptomics data. Our proposed Bayesian hierarchical model first links the network structures within each platform using a Markov random field prior to relate edge selection across sample groups, and then links the network similarity parameters across platforms. This enables joint estimation in a flexible manner, as we make no assumptions on the directionality of influence across the data types or the extent of network similarity across the sample groups and platforms. In addition, our model formulation allows the number of variables and number of subjects to differ across the data types, and only requires that we have data for the same set of groups.  We illustrate the proposed approach through both simulation studies and  an application to gene expression levels and metabolite abundances on subjects with varying severity levels of Chronic Obstructive Pulmonary Disease (COPD). }{Data integration; Gaussian graphical model; Bayesian inference; Markov random field prior; spike and slab prior; chronic obstructive pulmonary disease (COPD)}
\end{abstract}

\section{Introduction}
\label{sec:intro}
Gaussian graphical models, which describe the dependence relations among a set of random variables, have been widely applied to estimate biological networks on the basis of high-throughput data. When all samples are collected under similar conditions or reflect a single type of disease, methods such as the graphical lasso \citep{Meinshausen, Yuan2007, Friedman2008} or Bayesian network inference approaches \citep{Roverato2002, Wang2012} can be applied to infer a sparse network. In many studies, however, such as the COPDGene study \citep{regan} of this paper, described below, samples are obtained for different subtypes or disease, varying experimental settings, or other heterogeneous conditions. In this setting, applying standard graphical model inference approaches to the pooled data across conditions will lead to spurious findings, while separate estimation for each subgroup reduces statistical power. The challenge becomes even more formidable when multiple data types (or platforms) are under consideration, specifically gene expression levels and metabolite abundances in the COPDGene study, measured on multiple subjects.
In this case, pooling the data is not appropriate, as it ignores the fact that direct connections between variables of different data types may not be sensible. Nonetheless, analyzing data from each platform separately ignores potential commonalities, for example, that subjects with more advanced disease may have more extensive disruption of functional mechanisms across data types. The need for statistical methods to address these questions is particularly pressing given the increasing number of studies investing in comprehensive profiling of subjects across multiple data platforms. Our proposed statistical method enables joint inference of networks across sample groups and data types, providing accurate characterization of complex disease mechanisms which can be used to develop targeted treatment approaches.

Recently, methods have been proposed to estimate multiple networks on a common set of variables. Early work includes approaches that encourage either common edge selection or precision matrix similarity by penalizing cross-group differences \citep{Guo2011, Danaher2012, Zhu2014, Cai2016}. These methods use a single penalty parameter to control network similarity across all groups.  \cite{Hao2018} have extended the approach to simultaneously infer graph clustering via an additional penalty on the estimated cluster mean. In contrast, more recent proposals encourage network similarity in a more tailored manner, assuming that the networks for each sample group are related within a tree structure \citep{Oates2014}, or, more generally, within an undirected weighted graph \citep{Saegusa2016, Ma2016}. These methods require that the cross-group relations are known a priori or inferred in a preliminary step. More flexible approaches that employ Bayesian frameworks to simultaneously learn the networks for each group and their similarity have been proposed in \cite{peterson} and \cite{shaddox}. 

In this work, we develop a graphical modeling framework which enables the joint inference of network structures when there is heterogeneity among both sets of subjects (i.e., at different stages of a disease) and sets of variables (i.e., types of data or platforms).  Our proposed Bayesian hierarchical model first links the network structures within each platform using a Markov random field prior to relate edge selection across the sample subgroups, and then links the measures of cross-group similarity across platforms. This is a flexible modeling approach, which allows the number of variables and number of subjects to differ across the data types, and only requires that we have data for the same set of subgroups.  Consider for example, the gene expression and metabolite abundances measured on healthy controls and on moderate and severe COPD subjects of our case study. These two platforms measure different aspects of the same biological pathway.  Small compounds and metabolites are measured by the LC/MS platform, while gene expression levels of enzymes and proteins are measured by the microarray platform. Also, alterations in the pathway affect different components (metabolites or enzymes) of the pathway. In this type of scenario, we can expect data between the two platforms to be related. Our modeling framework is concerned with two types of network similarity-within and between platforms.  Within each platform, we assess how similar subgroups are in terms of their graph structure.  This results in a super-graph for each platform expressing whether two subgroups are similar, i.e., connected, within each platform.  We then assess whether or not these super graphs are similar between platforms.  This approach enables the joint estimation of the biological networks in a flexible manner, as we make no assumptions on the directionality of influence across the data types, nor on the extent of network similarity across the sample groups and platforms. In this regards, our approach differs from many of the existing methods for integrative analysis, that typically model the association between different types of observed random variables assuming a direction of influence among the data types, see for example \cite{wang2013} and \cite{cassese} for the use of multi-component hierarchical models, \cite{Chen2015} for mixed graphical models, and \cite{lin} for a multi-layered Gaussian graphical model where directed edges are allowed across layers of each data type.  Instead, we infer measures of relative similarity based on the data, which provide valuable insight into the extent of network relatedness across sample groups and data types. 

The paper is organized as follows. We present the motivating Chronic Obstructive Pulmonary Disease (COPD) case study in Section \ref{sec:data}. In Section \ref{sec:method}, we describe the proposed Bayesian model and procedures for posterior inference. We return to the COPD data set in Section \ref{sec:COPD2}, where we apply our proposed method to infer metabolic and gene co-expression networks for varying disease stages. Section \ref{sec:sim} provides simulations studies illustrating the performance of the proposed method against competing approaches.  Finally, we conclude with a discussion in Section \ref{sec:disc}.
 
\section{The COPDGene Study}\label{sec:data}

Our work has been motivated, in particular, by a collaborative study aimed at understanding how cellular metabolic and gene expression networks are disrupted by COPD, the 3rd leading cause of death in the United States \citep{NCHS}, and one of the top causes of hospitalization. While smoking is the primary risk factor for COPD, only 20\% of smokers will ever develop the disease. There is a poor understanding of risk factors accounting for disease susceptibility, as well as the underlying pathogenic mechanisms resulting in airway inflammation and emphysema. Understanding the genetic, clinical, and molecular factors that determine why some smokers develop COPD is the primary motivation of the NIH funded multicenter observational study, COPDGene, which has over 10,000 participants and includes extensive molecular profiling using transcriptomics, metabolomics, and proteomics. For this study, subjects 45-80 years old with at least a 10 pack-year history of smoking were recruited and biomarker measurements were attained from blood \citep{regan}.  There is a high degree of heterogeneity in the patient population, which includes subjects from various clinical stages, defined using the Global Initiative for Chronic Obstructive Lung Disease (GOLD) staging criteria. We apportioned subjects according to GOLD COPD stages and model resulting networks for a control group (GOLD stage = 0), a mild or moderate group (GOLD Stage = 1 or 2), and a severe group (GOLD Stage = 3 or 4). Here we focus in particular on a subset of subjects for whom gene expression levels or metabolite abundances are available. For the gene platform, this apportionment resulted in a control group (GOLD Stage  = 0) of 42 subjects, a mild or moderate group (GOLD Stage = 1 or 2) of 42 subjects, and a severe group (GOLD Stage = 3 or 4) of 42 subjects. For the metabolite platform, the control group again had 42 subjects, while the moderate and severe group had respectively 45 and 44 subjects.  Ten subjects had GOLD Stage = -1, indicating that although they had abnormal lung function, they didn't satisfy the clinical criteria for COPD.  These subjects were therefore excluded from the analysis. This data set illustrates the need for our proposed method, which can be used to analyze the multi-platform data across the heterogenous patient groups in a coherent and integrative fashion.  In summary, this paper is concerned with the analysis of data measured on two platforms, genes and metabolites, for three subgroups of subjects classified by COPD GOLD stage.

\section{Proposed Method} \label{sec:method}
\begin{figure}
\centerline{
\includegraphics[width=70mm]{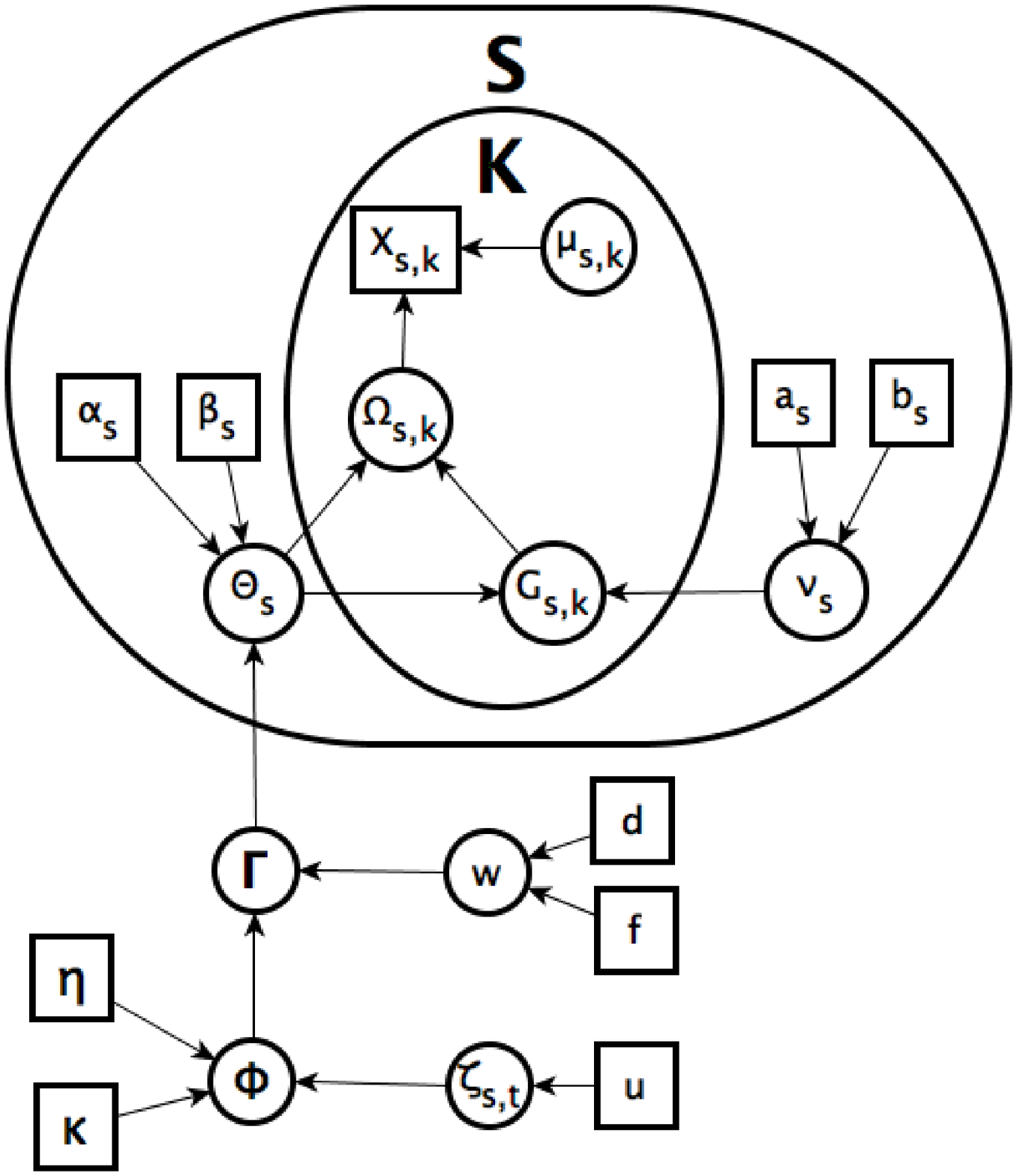}
\includegraphics[width=80mm]{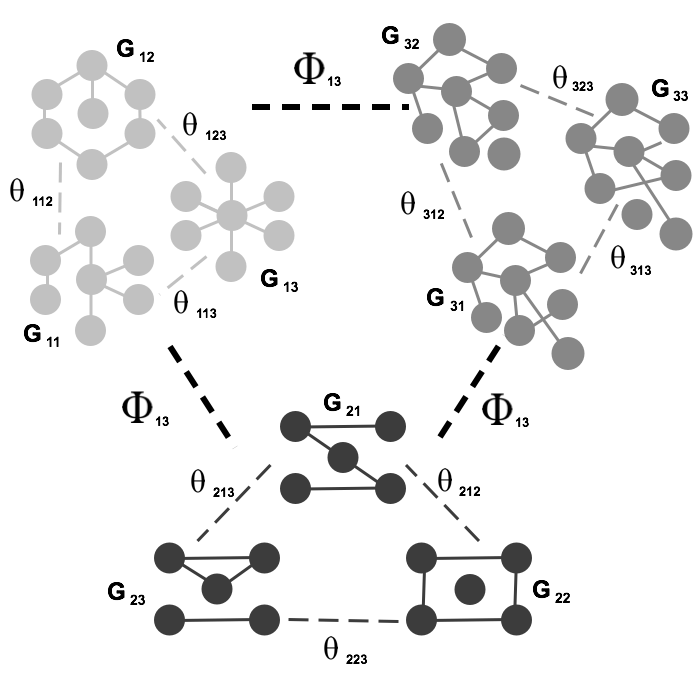}
}
\caption{{\it Left:} Graphical model representation of the proposed model, illustrating variables, parameters, and hyper parameters for each of the $K$ groups and $S$ platforms. {\it Right:} A graphical illustration with $K=3$ subgroups and $S=3$ platforms.}
\label{fig1graph}
\end{figure}

In this section, we provide details on the proposed method, including the likelihood, prior formulation, and procedures for posterior inference. Graphical representations are provided in Figure \ref{fig1graph}.

\subsection{Likelihood}
Suppose we observe data on $S$ data types and $K$ subgroups. In our COPDGene case study, we have $S=2$ and $K=3$.  For each subgroup and each platform, let $\mathbf{X}_{sk}$ be the $n_{sk}\times p_s$ data matrix, with $k=1, \ldots, K$ indexing the subgroup, $s=1, \ldots, S$ indexing the platform type, $n_{sk}$ the sample size for subgroup $k$ from platform $s$, and $p_s$ the total number of observed variables for platform $s$.  Assuming that the samples are independent and identically distributed within each of the $K$ subgroups and $S$ platforms, we can write the likelihood for subject $l$ in subgroup $k$ and platform $s$ as the multivariate normal distribution
\begin{equation} \label{likelihood}
X_{skl}\sim \mathcal{N}(\bm{\mu}_{sk}, \mathbf{\Omega}_{sk}^{-1}),\hspace{.3cm}l=1, \ldots, n_{sk},
\end{equation}
where the mean vector $\bm{\mu}_{sk}\in \mathbb{R}^{p_s}$ and precision matrix $\mathbf{\Omega}_{sk}=\mathbf{\Sigma}_{sk}^{-1}$ are specific to subgroup $k$ and platform $s$. For simplicity, we column-center the data for each subgroup and therefore assume ${\bm{\mu}_{sk} }= 0$. We note that $\mathbf{\Omega}_{sk}$ is constrained to the space $M^+$ of $p_s \times p_s$ positive-definite symmetric matrices. We denote the entry in the $i$th row and $j$th column of $\mathbf{\Omega}_{sk}$ as $\omega_{skij}$.

\subsection{Prior formulation}
The patterns of zeros in the precision matrices $\mathbf{\Omega}_{sk}$ correspond to undirected graphs among the variables. Specifically, $\omega_{skij}=0$ if and only if the corresponding edge $(i,j)$ is missing in the conditional dependence graph for subgroup $k$ from platform $s$. The goal of our modeling formulation is to infer a sparse version the precision matrices $\mathbf{\Omega}_{sk}$  in a manner that links inference across platforms. 

The graph for each platform $s$ and subgroup $k$ can be defined by a set of vertices $V=\{1, \ldots, p_s\}$ and edges $E\in V\times V$, and may be expressed as a symmetric binary matrix $\mathbf{G}_{sk}$, where each off-diagonal element $g_{skij}$ denotes the inclusion of edge $(i,j)$. Our proposed model first links the edge inclusion indicators across sample subgroups within each platform, and then links platforms based on the dependences across subgroups within each platform. We now describe in detail the components of our prior. 

\subsubsection{Mixture prior on precision matrix elements}

We rely on the mixture prior proposed in \cite{wang2015} to infer a sparse version of $\mathbf{\Omega}_{sk}$. This prior is attractive as it allows direct modeling of the latent graph $\mathbf{G}_{sk}$ and is computationally scalable. Mathematically, it can be expressed as a product of $p_s(p_s-1)/2$ normal mixture densities on the off-diagonal elements, and $p_s$ exponential densities on the diagonal elements, normalized to have total probability of one. This is equivalent to a hierarchical model
\begin{align}
\label{precision}
 p(\mathbf{\Omega}_{sk}| \mathbf{G}_{sk}, \nu_0, \nu_1, \lambda) &\propto
 \prod\limits_{i<j}\mathcal{N}(\omega_{skij}| 0, \nu_{g_{skij}}^2)\prod_i {\textnormal{Exp}}(\omega_{skii}|\frac{\lambda}{2}) \notag \\
 p(\mathbf{G}_{sk}| \nu_0, \nu_1, \pi, \lambda) &\propto
 \prod_{i<j} \big\{\pi^{g_{skij}} (1-\pi)^{1-g_{skij}} \big\}, 
\end{align}
where $\nu_{g_{skij}}=\nu_1$ if edge $(i,j)$ is present in graph $\mathbf{G}_{sk}$ and $\nu_{g_{skij}}=\nu_0$ otherwise, with $\nu_0$ ($< \nu_1$) being set to a small number.  The two component normal mixture model has been shown to be a successful prior in the context of variable selection, which in our case is equivalent to edge selection, and the choice of hyperparameters $\nu_0$ and $\nu_1$ has been closely studied by \cite{george1993}.  If for example, $\nu_0$ is chosen to be small, the event $g_{i,j}=0$ indicates that the edge $\omega_{i,j}$ comes from the $N(0,\nu_0^2)$ or diffuse component of the mixture, and consequently $\omega_{i,j}$ is closer to zero and can be estimated as zero.  In contrast, if $\nu_1$ is chosen to be large, the event $g_{i,j}=1$ means $\omega_{i,j}$ comes from the other component $N(0, \nu_1^2)$ and $\omega_{i,j}$ can then be thought of as substantially different from zero.

\subsubsection{Markov Random Field priors for linked network inference}

Markov Random Field (MRF) priors \citep{besag} have been successfully employed to capture network structure among the variables in Bayesian variable selection modeling frameworks \citep{lizhang, stingo} and more recently to link the selection of edges across multiple networks \citep{peterson, shaddox}. Here we build upon this line of work and utilize MRF priors both to link edge selection across networks within a platform, and to link the network similarity parameters across platforms. 

\noindent
\textit{Prior linking networks within each platform:} 
\noindent
Let $\mathbf{g}_{skij}=\{g_{s1ij},\ldots, g_{sKij}\}^T$ represent the vector of binary inclusion indicators of edge $(i,j)$ across the $K$ graphs for platform $s$. We define a MRF prior on this vector of binary inclusion indicators, linking edge selection across  networks within a platform as 
\begin{equation}
\label{subgroupMRF}
p(\mathbf{g}_{skij}| \nu_{sij}, \mathbf{\Theta}_s)=\frac{\exp(\nu_{sij}{\bf{1}}^T{\mathbf{g}_{skij}}+{\mathbf{g}_{skij}}^T\mathbf{\Theta}_s{\mathbf{g}_{skij}})}{C(\nu_{sij}, \mathbf{\Theta}_s)},
\end{equation}
with $\nu_{sij}$ a sparsity parameter and $\mathbf{\Theta}_s$ a $K\times K$ symmetric matrix characterizing pairwise relatedness across sample subgroups. The diagonal elements of $\mathbf{\Theta}_s$ are constrained to be 0, while the off-diagonal elements $\theta_{skm}$ drive the {\it within platform similarity} and link the edge selection between sample subgroups $k$ and $m$, such that a larger magnitude represents increased preference for shared similar structure between those two subgroups.  In our experience, these entries can be interpreted on a relative rather than an absolute scale, as magnitude can vary depending on hyperparameter settings, although ordering is generally preserved.  Additionally, the vector of binary inclusion indicators allows easy interpretation of the off-diagonal elements of $\theta_{skm}$ as regression coefficients of a probit model.  If we introduce the notation $\nu_s=\{\nu_{sij}|1\leq i<j\leq p_s\}$, then we can write the joint prior across graphs $\mathbf{G}_{sk}$ for platform $s$ as the product of the densities for each edge as 
\begin{equation}
\label{jointGraphPrior}
P(\mathbf{G}_{s1}, \ldots, \mathbf{G}_{sK}|\nu_s, \mathbf{\Theta}_s)=\prod\limits_{i<j} p(\mathbf{g}_{skij}|\nu_{sij}, \mathbf{\Theta}_s).
\end{equation}

Imposing sparsity on the matrix $\mathbf{\Theta}_s$ results in a ``super-graph" describing relatedness of the networks across the sample subgroups, with zero entries indicating that the networks are sufficiently different that edge selection should not be shared. This is achieved assuming a spike-and-slab prior on the off-diagonal entries of $\mathbf{\Theta}_s$, with a Gamma as the slab, since only positive values for $\theta_{skm}$ are sensible, 
\begin{equation}
\label{thetaPrior}
P(\theta_{skm}| \gamma_{skm})=(1-\gamma_{skm})\delta_0 +\gamma_{skm}\frac{\beta^{\alpha}}{\Gamma({\alpha})}\theta_{skm}^{\alpha-1}e^{-\beta\theta_{skm}},
\end{equation}
where $\Gamma(\cdot)$ represents the gamma function, $\alpha$ and $\beta$ are fixed hyperparameters, and the latent indicator variable $\gamma_{skm}$ indicates the event that the network for subgroup $k$ is related to subgroup $m$ on platform $s$.  The joint prior on the off-diagonal entries is the product of the marginal densities 
\begin{equation}
\label{jointTheta}
p(\mathbf{\Theta}_s|\bm{\gamma}_s)=\prod\limits_{k<m}p(\theta_{skm}|\gamma_{skm}).
\end{equation}
This prior construction allows sharing of information between subgroups when appropriate, without forcing similarity in cases where the networks are actually different. Additionally, we specify a prior on the sparsity parameter $\nu_{sij}$ as
\begin{equation}
\label{prior_vu}
P(\nu_{sij})=\frac{1}{\beta(a,b)}\frac{e^{a\nu_{sij}}}{(1+e^{\nu_{sij}})^{a+b}},
\end{equation}
where $\beta(\cdot)$ denotes the Beta function, and $a$ and $b$ are fixed hyperparameters.  Platform specific hyperparameters may be chosen in cases where sparsity is known to be different from one platform to another. 

\noindent
\textit{Prior linking cross-group relations across platforms:} 
To link networks at the platform level, we model the overall relationship between each pair of platforms based on the dependencies across subgroups within each platform. This is a flexible approach which allows the number of variables and number of subjects to differ across the platforms, and only requires that we have data for the same set of subgroups. Specifically, we construct an MRF prior on the vector of binary indicators for network relatedness between subgroups $k$ and $m$ across all platforms, $\bm{\gamma}_{km}=\{\gamma_{1km}, \ldots, \gamma_{Skm}\}^T$, as 
\begin{equation}
\label{platformMRF}
p(\bm{\gamma}_{km} | w_{km}, \mathbf{\Phi})=C(w_{km}, \mathbf{\Phi})^{-1}\exp(w_{km}{\bf{1}}^T \bm{\gamma}_{km}+\bm{\gamma}_{km}^T\mathbf{\Phi} \bm{\gamma}_{km}),
\end{equation}
with $w_{km}$ capturing the sparsity of the vector $\bm{\gamma}_{km}$ and $\mathbf{\Phi}$ a $S\times S$ symmetric matrix denoting pairwise similarity across platforms, in a similar manner to the matrices $\mathbf{\Theta}_s$ described previously.   The off- diagonal elements of $\mathbf{\Phi} $ drive the {\it{ between platform similarity}}, a non-zero $\phi_{st} $ indicates platforms $s$ and $t$ have similar super graphs $\mathbf{\Theta}_s$ and $\mathbf{\Theta}_t$. As above, we place a spike-and-slab prior on the entries of $\mathbf{\Phi}$,
\begin{equation}
\label{phiPrior}
p(\phi_{st}|\zeta_{st})=(1-\zeta_{st})\delta_1+\zeta_{st} \frac{\kappa^\eta}{\Gamma(\eta)}\phi_{st}^{\eta-1}e^{-\kappa\phi_{st}},
\end{equation}
with $\kappa$ and $\eta$ fixed hyperparameters, and $\zeta_{st}$ a latent binary variable which indicates that platforms $s$ and $t$ have related cross-group dependencies.  Off-diagonal entries $\phi_{st}$ in the symmetric $S\times S$ matrix signify the magnitude of pairwise relatedness across platforms, modeling the relations across different platforms as learned from the data in an innovative and versatile manner.  We then place independent Bernoulli$(u)$ priors on the latent indicators $\zeta_{st}$, with $u$ a fixed hyperparameter $\in [0,1]$, and specify a prior on $w_{km}$ similarly to (\ref{prior_vu}) with hyperparameters $d$ and $f$ to complete the model.

\subsection{Posterior inference}
Let ${\bf{\Psi}}=\{ {\mathbf{ \Omega}_{sk}},{\mathbf{ G}_{sk}},{\mathbf{ \Theta}_{s}}, \nu_{sij},{\bm{ \gamma}_{s}}, w_{km},{\mathbf{ \Phi}},{\bm{ \zeta}}\}$ denote the set of all parameters and ${\bf{X}}$ denote the observed data for all sample subgroups and all platforms.  We can write the joint posterior as 
\begin{equation}
\begin{array}{rl}
p({\bf{\Psi}}| {\bf{X}})\propto&\prod\limits_{s=1}^S\bigg\{\prod\limits_{k=1}^K \biggr[p({\mathbf{X}_{sk}|  \mathbf{\Omega}_{sk}})\cdot p({\mathbf{\Omega}_{sk}}| \mathbf{G}_{sk})\biggr]\\ 
&\times\prod\limits_{1\leq i<j\leq p_s}\biggr[P({\bf{g}}_{sij}|\nu_{sij}, \mathbf{\Theta}_s)\cdot p(\nu_{sij})\biggr]\cdot p({\mathbf{\Theta}_s}| \bm{\gamma}_{s})\bigg\}\\ 
&\times\prod\limits_{k<m}\biggr[p(\bm{\gamma}_{km}| w_{km}, \mathbf{\Phi})p(w_{km})\biggr]\cdot p(\mathbf{\Phi}| \bm{\zeta})\cdot p(\bm{\zeta}).
\end{array}
\end{equation}
As this distribution is analytically intractable, we construct a Markov chain Monte Carlo (MCMC) sampler to obtain a posterior sample of the parameters of interest.

\subsubsection{MCMC sampling scheme}
Our MCMC scheme includes a block Gibbs sampler to sample the precision matrix $\mathbf{\Omega}_{sk}$ and graph $\mathbf{G}_{sk}$ for each platform $s$ and subgroup $k$.  Then we sample the graph similarity parameters $\mathbf{\Theta}_s$ and $\bm{\gamma}_s$ for each platform using a Metropolis-Hastings method that is equivalent to a reversible jump and incorporates between-model and within-model moves. Next, we use Metropolis-Hastings steps to sample the edge-specific sparsity parameters $\nu_{sij}$ and the cross-subgroup relation sparsity parameters $w_{km}$ from their respective posterior conditional distributions. Lastly, we update the cross-platform parameters $\mathbf{\Phi}$ and $\bm{\zeta}$  using a Metropolis-Hastings method similarly to the one used to update $\mathbf{\Theta}_s$ and $\bm{\gamma}_s$.
A detailed description of the MCMC algorithm is provided in the Supplementary Material.

\subsubsection{Model selection}
There are various approaches for making inference on the graph structures based on the MCMC output.  One approach is to use the maximum a posteriori (MAP) estimate, which represents the mode of the posterior distribution for each graph.  However, this approach is generally not preferred in the context of large networks since the space of possible graphs is large and we may only visit a particular graph a few times during the MCMC.  We then rely on a more practical approach for model selection, and estimate the marginal posterior probability (MPP) of inclusion for each edge $g_{skij}$, which we calculate as the proportion of MCMC iterations, after burn-in, where edge $(i,j)$ was included in graph $\mathbf{G}_{sk}$.  Final inference is performed by selecting edges according to the median model (i.e., with MPP$>0.5$) for inclusion in our posterior selected graphs \citep{barbieri2004}.

\section{Case Study on COPD Disease Severity} \label{sec:COPD2}
We are interested in studying the reshaping of gene and metabolite networks as disease stage worsens.  Our ultimate goal is to be able to map the underlying molecular causes of disease progression and to determine whether biological platforms describe the same mechanisms. 

Gene expression levels were measured from peripheral blood mononuclear cells (PBMCs) using the Affymetrix Human Genome U133 Plus 2.0 Array \citep{k1}, and plasma metabolite abundances were generated from Liquid Chromotography/Mass Spectrometry \citep{bowler2014}.  Candidate pathways were selected as follows.  Differently expressed genes and differently abundant metabolites were identified for airflow obstruction (FEV1pp forced expiratory volume in 1 second percent predicted) correcting for age, sex, body mass index, and current smoking status. KEGG Pathways \citep{kanehisa} that showed enrichment of the significant genes and metabolites were then prioritized. Top candidate pathways may play a role in the response to cigarette smoke exposure and are interesting candidates for more detailed exploration in emphysema.  

Below we report results on one of the top candidate pathways we analyzed, Regulation of autophagy (RegAuto).  Results on a second candidate pathway, Fc$\gamma$R-mediated phagocytosis (Fc$\gamma$R), can be found in the Supplementary Material. Expression levels were measured for 28 (RegAuto) and 104 (Fc$\gamma$R) probesets. These were collapsed to  20 (RegAuto), and 58 (Fc$\gamma$R) unique genes by selecting, for each gene, the probeset with the strongest association to emphysema.  Metabolite data was matched to lipid and aqueous annotation files in order to extract KEGG IDs for each sample.  After subsetting to the RegAuto and Fc$\gamma$R pathways, we were left with 117 (RegAuto) and 60 (Fc$\gamma$R) measurements, but numerous instances of duplicate KEGG IDs.  To reduce redundancy and exclude highly correlated covariates, we carried out an iterative principal component analysis procedure to select a subset of less correlated variables for analysis.  This procedure is outlined in the Supplementary Material and an example code is provided online. After this procedure,  21 (RegAuto) and 23 (Fcy$\gamma$R) metabolites were left for analysis.
    
\subsection{Hyperparameter settings}
The application of our model requires the specification of several hyperparameters. Here we describe the specification we used to obtain the results reported below and refer to the simulation study for more insights and sensitivity analyses. For prior (\ref{precision}) on the precision matrix elements, hyperparameters were specified as $\nu_0=.02$ and $\nu_1 = 1$ according to published guidelines given in \cite{wang2015}.  As for the prior (\ref{thetaPrior}) on the off-diagonal entries of $\Theta_S$ linking sample subgroups within a platform, we specified the slab portion of the mixture prior as a Gamma$(\alpha, \beta)$ with $\alpha=1$ and $\beta = 9$ for both platforms.  This resulted in a prior with mean approximately equal to .1 and $P(\theta_{km} \leq 1) = .99$, which avoids assigning high values to the off diagonal entries of $\Theta_s$.  For the prior (\ref{prior_vu}) on the sparsity parameter $\nu_{sij}$ of the MRF prior linking networks within each platform, we specified $a = 1$ and $b = 7$ resulting in a prior probability of edge inclusion around .125.  The similarly specified prior on sparsity parameter $w_{km}$ in the MRF prior (\ref{platformMRF}) linking cross-subgroup relations across platforms was specified as $d=1$ and $f=19$, for all subgroup pairs $k,m$, resulting in approximately $5\%$ prior probability of subgroup relatedness.  The mixture prior (\ref{phiPrior}) on the off-diagonal entries of $\Phi$ was specified as Gamma$(\eta, \kappa)$ with $\eta = 4$ and $\kappa = 5$, resulting in a prior mean of $.4$ and $P(\phi_{st} \leq 1) = .96$, avoiding assigning high values to the off diagonal entries of $\Phi$.  Lastly, the hyperparameter $u$ in the Bernoulli prior on the indicators of platform similarity $\zeta_{st}$ was specified as $u=.1$. 
Sensitivity analyses reported in the Supplementary Material show that hyperparameter settings have minimal impact on graph learning performance as the inferred network remains fairly stable.  With certain settings, large changes may occur in the magnitude of relative similarity measures $\Theta_s$ and $\Phi$, however ordering is generally preserved. 
Results we report here and in the Supplementary Material were obtained by running two MCMC samplers for 10,000 burnin iterations followed by 30,000 iterations used for inference, with different starting points.  To verify convergence of the chains, we compared correlations of resulting MPPs from the two chains.  Those were in the range $(.9357, 1.000)$, for Pearson correlations. Final results were obtained by pooling together the output of the two chains. 

\subsection{Results}
Estimated graphs for control, moderate, and severe subgroups, for the RegAuto pathway, obtained by selecting edges with MPPs greater than 0.5,  are shown in Figure \ref{fig_RegAutoMetabolite}, and those for the Fc$\gamma$R pathway are reported in the Supplementary Material. In these plots, obtained using the software {\it cytoscape} \citep{cytoscape}, the size of a node is drawn proportionally to the number of edges connecting that node to others in the same graph (i.e., the ``degree" of the node).  For the RegAuto pathway, relative network similarities across subgroups were estimated as

\[MPP(\Theta)^{Genes}_{RegAuto}= \left( \begin{array}{ccc}
\cdot & .9932 & .9741 \\
 &  \cdot& .9861 \\
 &  & \cdot \end{array} \right) \hspace{1cm}
MPP(\Theta)^{Metabolites}_{RegAuto} =  \left( \begin{array}{ccc}
 \cdot&  .9562& .9618 \\
 & \cdot & .9560 \\
&  & \cdot \end{array} \right)\] 

\noindent with relative similarity across platforms estimated as $MPP(\Phi)_{RegAuto} =.9685$.  These values indicate a preference for shared structure across platforms and sample subgroups. Histograms of posterior distributions of non-zero values of $\Theta_S$ and $\Phi$ are shown in the Supplementary Material.

Table \ref{t:one} indicates the total number of inferred pair interactions across the two pathways, together with the counts of pairs that exhibit evidence of disrupted interactions due to disease severity.  In the table, for each pathway, the three disease subgroups ordered from least to most severe are coded with 0's and 1's, with 1 indicating a high marginal posterior probability (MPP$>0.5$) of edge inclusion in the subgroup network.  For instance, 110 would indicate that the edge is present in the control and moderate subgroup, yet not in the severe subgroup.  That is, in the severe subgroup the MPP of edge inclusion falls below the threshold of 0.50.  Group codings of 100 and 110 indicate greater interaction in the control and group codings 011 and 001 indicate greater interaction in disease.  For the gene platform, counts for known protein-protein interactions are included in parentheses for the gene platform.  Biological General Repository for Interaction Datasets (BioGrids) v. 3.4.156 \citep{Chatr} was used to obtain protein-protein interactions and disease annotation information was gathered from \cite{stelzer}.  We observe $50-60\%$ disruption in total pairs of genes and metabolites, and different patterns of disruption for metabolites and gene interactions. For both metabolites and genes, there are a large number of connections in control subjects that are then disrupted in moderate/severe subjects. But for metabolites, there is also a relatively large number of metabolite connections in severe subjects that are not present in the moderate/control subjects, suggesting that parts of the metabolite pathway are activated as disease severity increases. These results also illustrate that while our method takes advantages of commonalities between the platforms, it can also highlight platform specific differences.

In order to gain further intuition on the properties of the estimated graphs, we calculated a number of graph metrics across all subgroups and pathways. Results on number of edges, global clustering coefficient, averaged betweenness centrality and count of hub nodes are reported in Table \ref{t:two2}. The global clustering coefficient of a graph is based on node triplets, i.e. 3 connected nodes, and is defined as the number of closed triplets divided by the total number of connected triplets. It measures the degree to which nodes in a graph tend to cluster together, with values closer to 1 if the graph is more modular i.e. it can be divided into clusters of highly connected nodes. Betweenness centrality quantifies the number of times a node acts as a bridge along the shortest path between two other nodes, as a measure of how important the node is in serving as a connector between other nodes in the graph.  

A close inspection of the estimated networks and our results suggests that, in general, estimated gene networks exhibited a trend of decreased connectivity or a large drop in connections as disease severity increased, while metabolite networks do not show such a trend. There may be several reasons why the network patterns are different between the genes and metabolites. One possible reason is that the same metabolites are present in other biological pathways that may be compensating for the changes due to disease. Another reason is that the plasma metabolomics may be reflecting activity in multiple organs, while the gene level data is primarily reflecting changes in gene expression more specifically in the blood. Additionally, results in Table \ref{t:two2} generally indicate higher global clustering coefficients and degree centrality measures for gene platforms than for metabolite platforms.  This suggests that gene networks are generally more clustered into denser subnetworks characterized by high connectivity within each pathway when compared to metabolite networks.  Additionally, interpreting degree centrality measures in the context of information flow within networks suggests that when disrupted, highly connected genes may impact network communication more than disrupted metabolite interactions.

\subsection{Hub node analysis}
Further analysis of the results was carried out on hub nodes, for both platforms, to validate findings with known protein-protein interactions and to examine disease related gene annotation.  
Hub listings were generated for each pathway and each platform to allow analysis of node connectivity and variations in connectivity as disease increased in severity. A summary of results is given in Table \ref{t:two2}, where hub node count for our application setting signifies the number of nodes per group with a degree $\geq 4$, or at least four connections.  As an example, for genes in the FC$\gamma$R pathway we find that the there are less connections per node, less hub nodes, and less connections in the Severe subjects compared to the Moderate/Control subjects suggesting that there is overall disruption for this pathway at the gene expression level.  In the supplementary materials, we provide biological background on specific genes, metabolites and connections in the estimated networks for the two pathways. 
 
\section{Simulation Studies} \label{sec:sim}
In this section we compare our proposed method with three recently proposed graphical model learning methods:  Fused Graphical Lasso, Group Graphical Lasso, and Hub Graphical Lasso. The first two methods are designed to learn the network structure of related subgroups \citep{Danaher2012}: The Fused Graphical Lasso encourages both shared structure and shared edge values, the Group Graphical Lasso encourages shared graph structures but not shared edge values. 
The Hub Graphical Lasso \citep{Mohan2014} encourages similarity across networks based on the presence or absence of highly-connected hub nodes. None of the competing methods encourages similarity across platforms. 

\subsection{Comparison study}
We investigate whether alternative methods can produce satisfactory results, in terms of network accuracy, in settings that mimic our COPD data (two platforms and three sampling subgroups).
We consider two set-ups for generating $p\times p$ adjacency and precision matrices, for each sampling group $k=1, 2, 3$:\\
	\indent $(1)$ Scale free networks: the probability that a given node has $e$ edges is proportional to $e^{-\alpha}$. We kept $\alpha=1$, the default setting as stated in the {\tt{igraph}} package \citep{igraph}, and simulated networks of the same size of pathways analyzed in the COPD case study ($p=40$ nodes).\\
	\indent $(2)$ AR(2) networks: the entries of the $p \times p$ precision matrix are defined as $\omega_{i,i}=1$ for $i=1, \ldots, p$, $\omega_{i,i+1}=\omega_{i+1,i}=0.5 $ for $i=1, \ldots, p-1$ and $\omega_{i,i+2}=\omega_{i+2, i}=0.4$ for $i=1, \ldots, p-2$. We simulated networks of larger size than pathways analyzed in the COPD case study ($p=80$ nodes).

As our model learns similarity between networks and does not enforce similarity unless supported by the data, current modeling allows for all patterns of similarity. In particular, from the preliminary adjacency matrices above, in our simulations we considered two settings of pairwise similarity across sampling groups for each platform: In setting one, for platform 1, Groups 1 and 2 were set up to be ``similar" while Group 3 was set up to be different. We generated ``similar" networks across all three groups for the second platform. Here, two groups are defined as ``similar" if the precision matrix of one group shares approximately $90\%$ of edges with the precision matrix of the other group. In setting 2, both platforms were set up to have different networks across all three subgroups. 

For scale free networks, to ensure that each generated precision matrix was positive definite, we used a similar approach to that of \cite{Danaher2012} where each off-diagonal element is divided by the sum of the off-diagonal elements in its row, and then the matrix is averaged with its transpose. Consequently, precision matrices generated via the scale free network method have lower signal, in terms of magnitude of the non-zero elements of the precision matrices, than the AR(2) networks; we simulated scale free networks of size $p=40$ and AR(2) networks of size $p=80$ to ensure a minimal signal strength. After all precision matrices were determined, data matrices $X_{s,k}$ of size $n=100$ for $k=1,2, 3$ and $s=1,2$, were generated from normal distributions $N(0, \Omega_{s,k}^{-1})$ and variables were standardized to have a standard deviation of one. We used the same hyperparameter setting used in the analysis of the COPD data, and ran our MCMC samplers for 10,000 burnin iterations followed by 30,000 iterations used for inference. Additional sensitivity analyses may be found in the Supplementary Material. Using a 2-core 1.7 GHz Intel core i7 processor with 8 GB memory, our code takes approximately 40 minutes to run 5000 iterations for a 2 platform scenario with 40 variables per platform. Alternative methods, such as the fused and group graphical lasso, are computationally more efficient, although grid searches and trials to determine optimized penalty parameters can be quite time consuming.

In Table \ref{tab:sim} we report network accuracy metrics averaged over 25 replicates; we considered the true positive rate (TPR), the false positive rate (FPR), the Matthews correlation coefficient (MCC), and area under the curve (AUC). Overall the proposed method performs comparatively well, and it is the only approach that controls the false positive rate across all scenarios. The differences in performances in favor of the proposed approach are particularly large in Setting Two. This is not surprising since the proposed approach is the only joint graph inference approach that learns from the data whether groups are related and, consequently, does not always enforce similarity across groups. Additionally, in the Supplementary Material we show a comparison of TPRs attained across methods for fixed FDRs, providing some evidence that our proposed method improves power with respect to methods that employ separate estimations for each subgroup.

\section{Conclusion} \label{sec:disc}
Motivated by a collaborative study on COPD progression, we have proposed a novel approach for joint multiple platform network analysis (here, genes and metabolites).  Our Bayesian approach uses computationally efficient priors on precision matrices and hierarchical MRF priors to link similarities across subgroups and platforms. Even though less scalable than alternative methods, a Bayesian framework makes use of all information in the data, sharing it across subgroups when appropriate, and enabling joint estimation in a very flexible manner, as we make no assumptions on the directionality of influence across the data types or on the extent of network similarity. In addition, our model formulation allows the numbers of variables and subjects to differ across data types. We have demonstrated improved performance over alternative approaches for multiple networks using simulated data.  On the COPDGene data, we have jointly inferred metabolite and gene networks across subgroups of disease stage, identifying notable interactions that illustrate disease progression and suggesting pathway compensation as a consequence to disease. These interactions pinpoint molecular targets for further study and provide potential therapy options.

\section*{Acknowledgements}
Work supported by NHLBI U01HL089897, U01HL089856, P20HL113445, and Butcher Foundation. Shaddox supported by NLM Training Program T15 LM007093; Peterson partially supported by NIH/NCI/P30CA016672.  COPDGene study (NCT00608764) supported by the COPD Foundation through contributions to an Industry Advisory Committee comprised of AstraZeneca, Boehringer-Ingelheim, GlaxoSmithKline, Novartis, Pfizer, Siemens and Sunovion. We also acknowledge support from NSF/DMS 1811568/1811445 and NSF/RTG 1547433 and thank Dessy Akinfenwa and Ami Sheth for help with the simulation study.

\section{Supplementary Material}
\label{sec6}

Supplementary material may be found online at \href{http://biostatistics.oxfordjournals.org} %
{http://biostatistics.oxfordjournals.org}.
Matlab code is available at \href{https://github.com/elinshaddox/MultiplePlatformBayesianNetworks} %
{https://github.com/elinshaddox/MultiplePlatformBayesianNetworks}.

\bibliographystyle{biorefs}
\bibliography{genesAndMetabolites_v7}

\begin{figure}
\centering
\includegraphics[width=40mm]{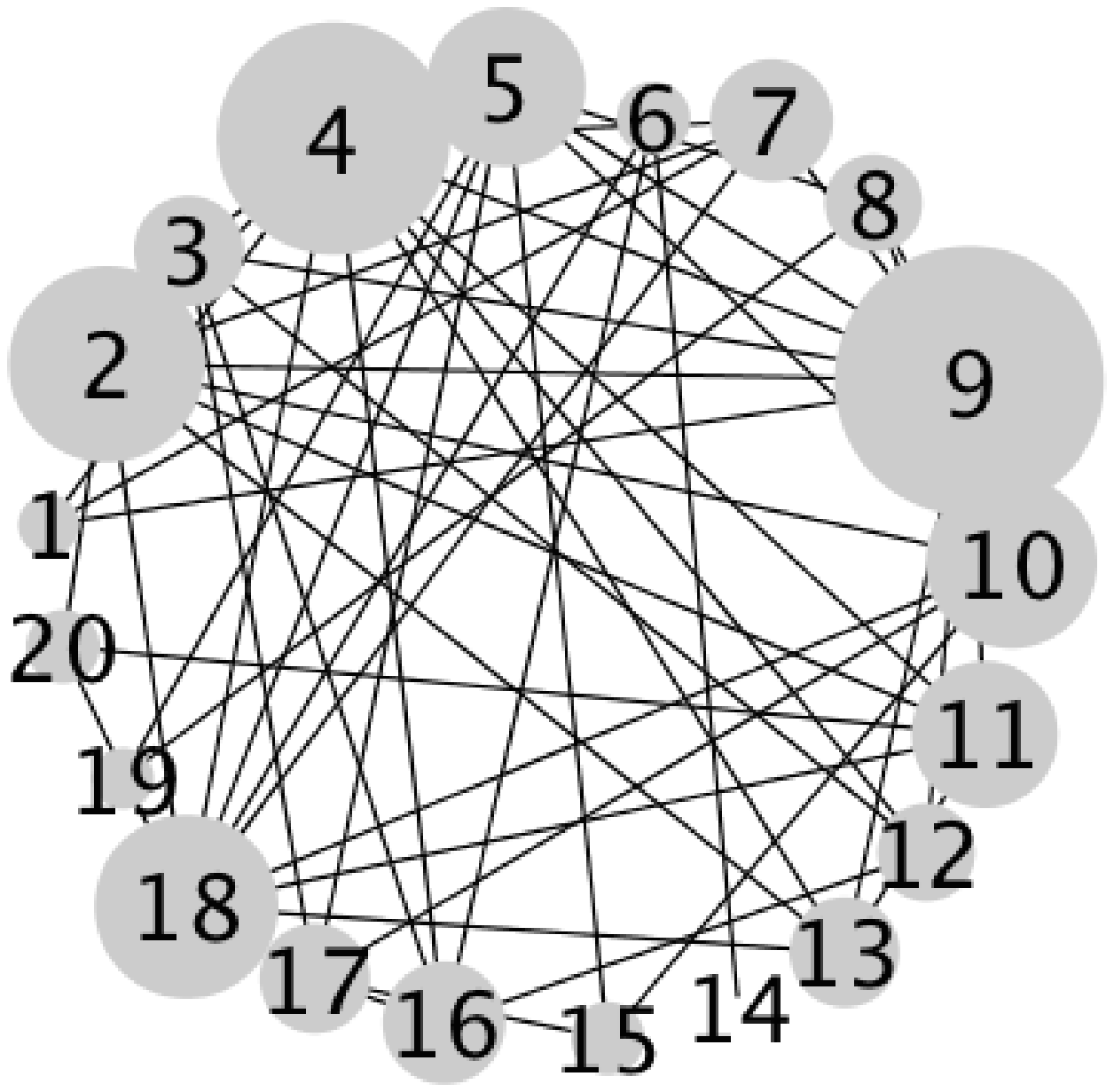}
\includegraphics[width=40mm]{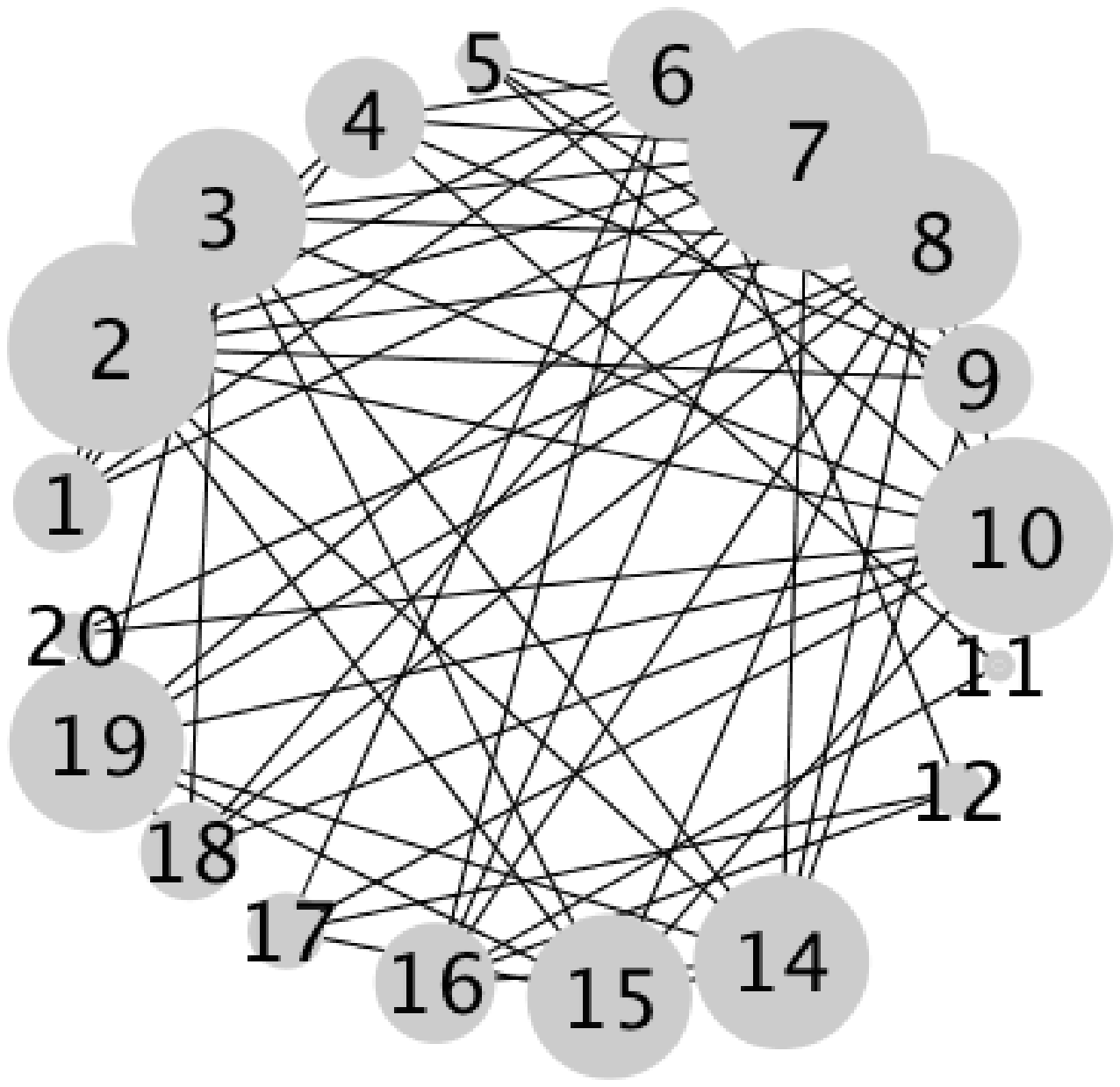}
\includegraphics[width=40mm]{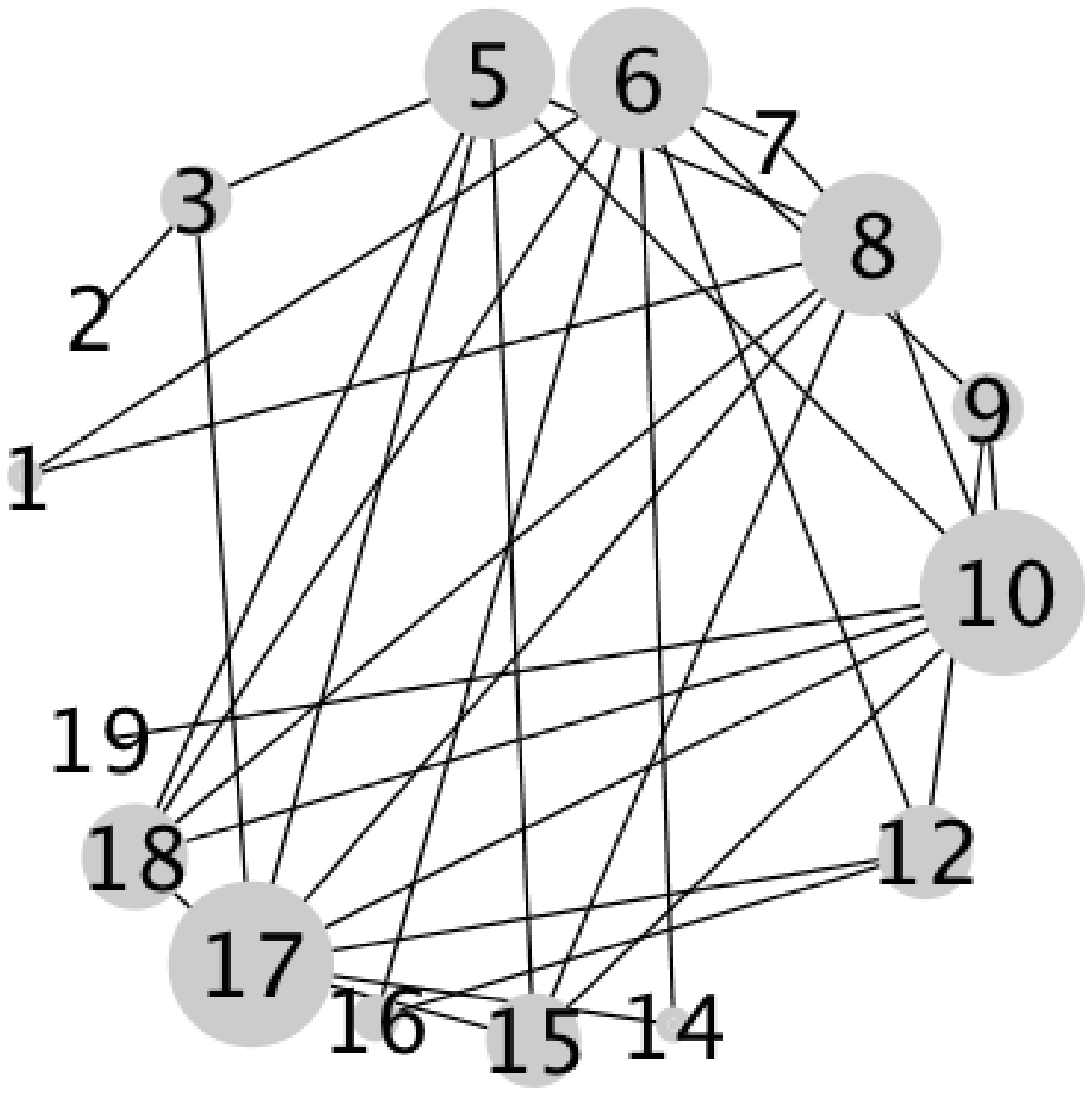}\\

\includegraphics[width=40mm]{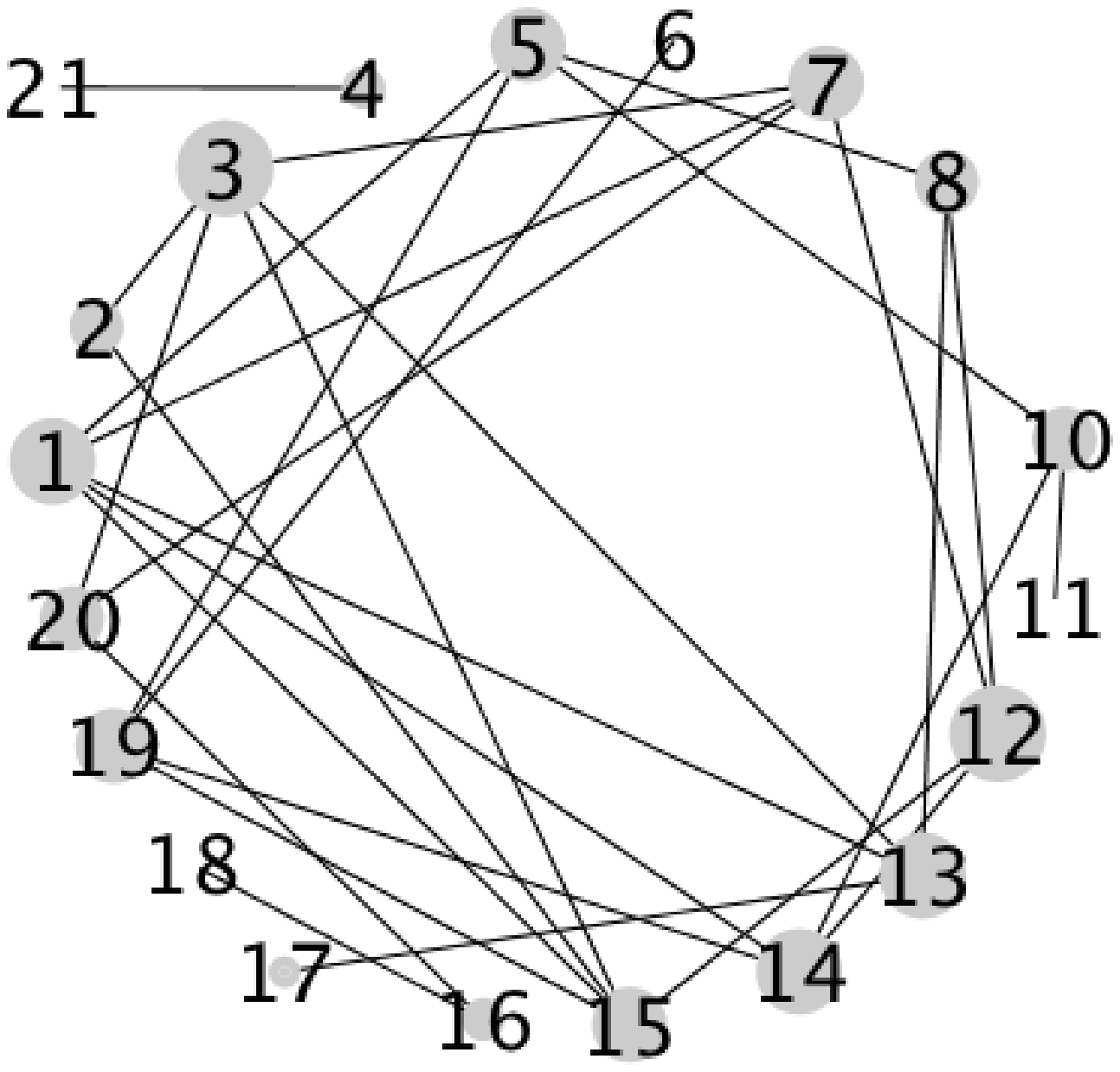}
\includegraphics[width=40mm]{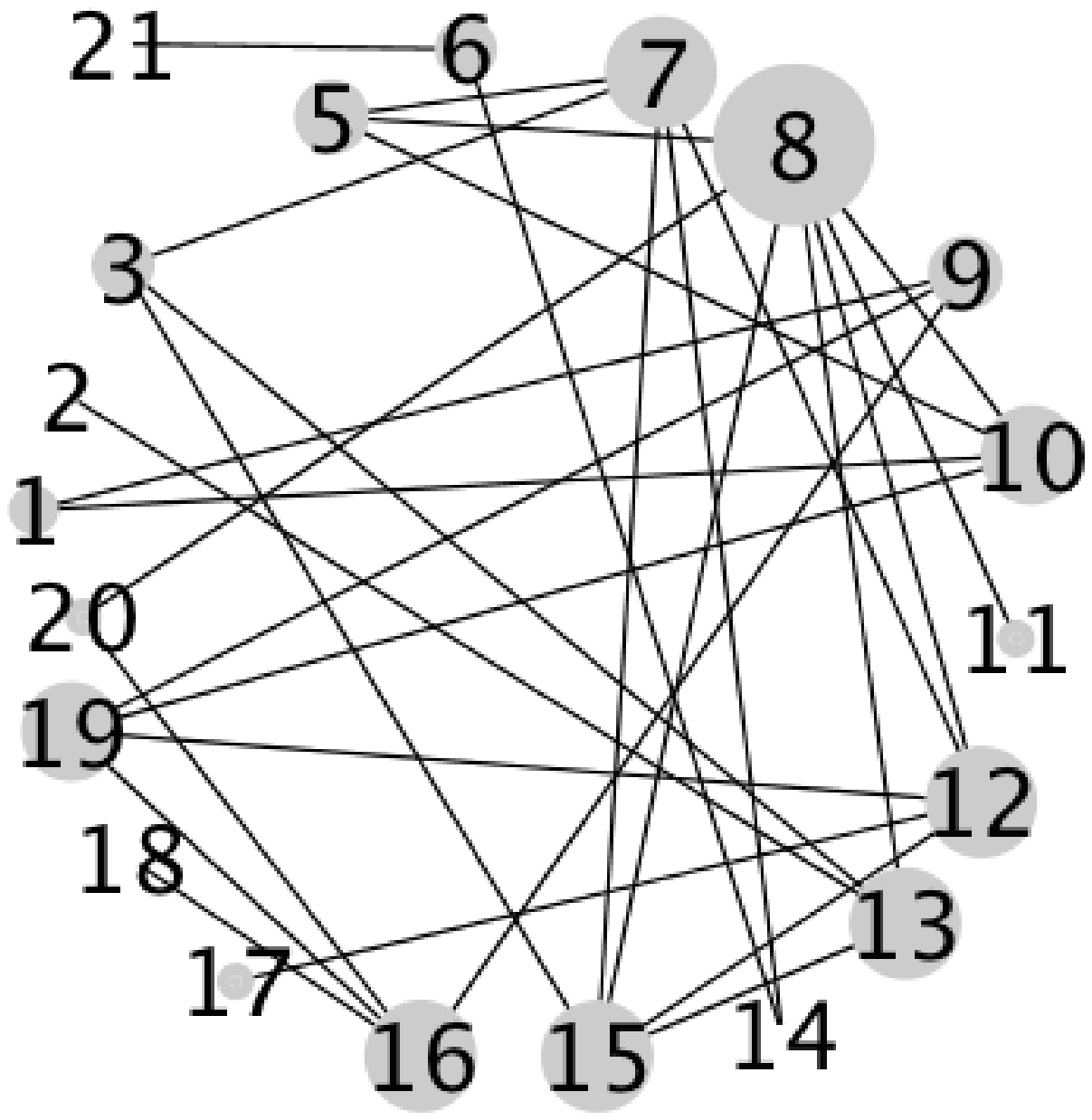}
\includegraphics[width=40mm]{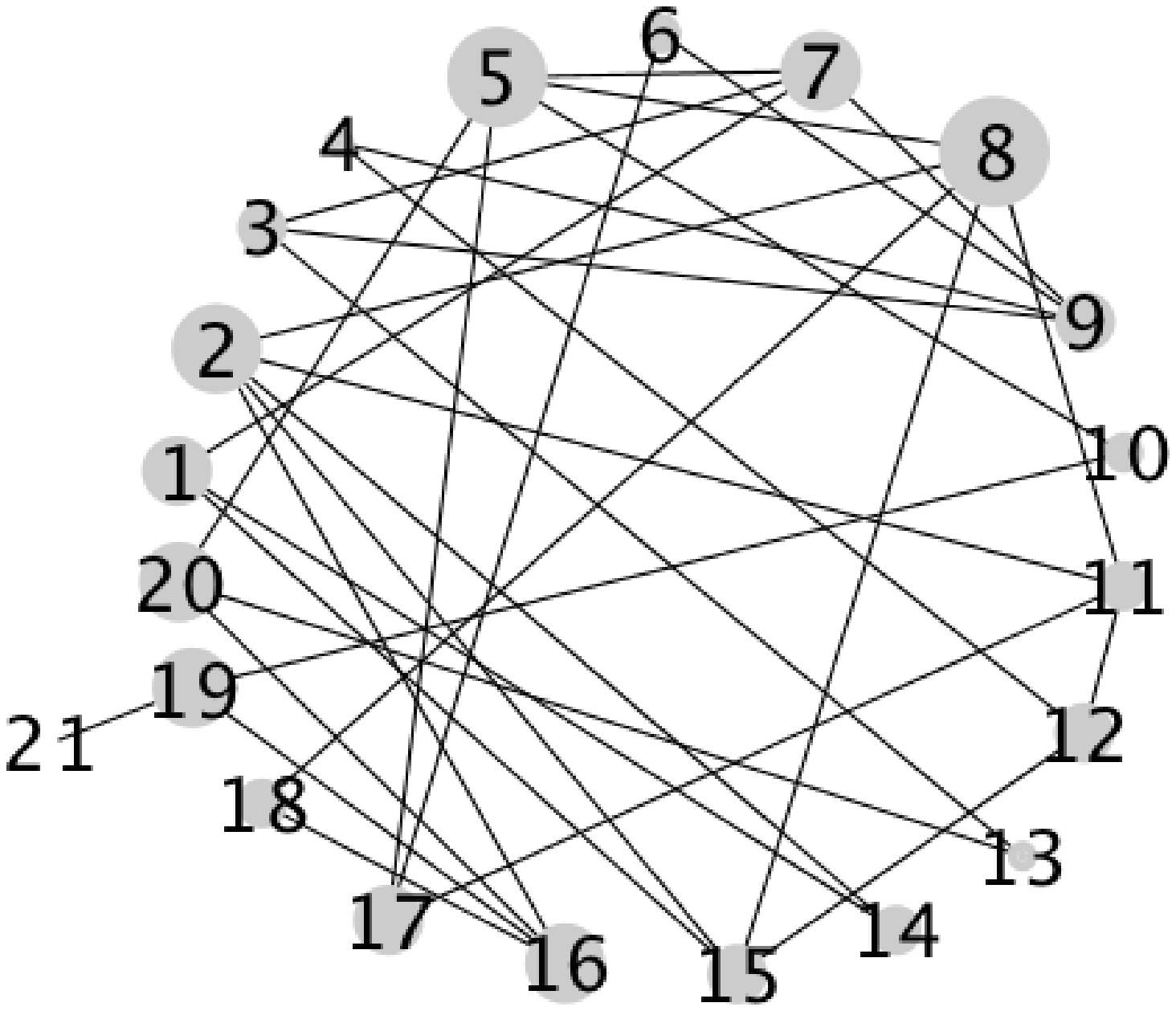}
\caption{Regulation of Autophagy (RegAuto) pathway, gene (top) and metabolite (bottom) platforms: Estimated graphs for control (left), moderate (middle), and severe (right) subgroups, obtained by selecting edges with MPPs greater than 0.5. The size of the nodes is proportional to their degree.}
\label{fig_RegAutoMetabolite}
\end{figure}

\begin{table}
\caption{Case study on COPD: numbers of total pairs of unique gene interactions and numbers of disease disrupted pairs based on disease severity.  Numbers in parentheses reflect the number of pairs with known protein protein interactions (PPIs).}
\label{t:one}
\begin{center}
\begin{tabular}{llcccccc}
\hline\hline
Pathway&Platform&Total Pairs &100& 110& 011 & 001 & Total Disrupted\\ \hline 
Fc$\gamma$R &Metabolites &73 & 17 & 5 & 3 & 18 & 43\\
Fc$\gamma$R &Genes & 656 (49) &151 (8) & 63 (7)  & 74 (3) & 63 (4) & 351 (22)\\
Reg Auto &Metabolites &66 & 14 & 4 & 5 & 17 & 40\\
Reg Auto &Genes & 101 (6) & 23 (2) & 13 & 7 & 8 & 51 (2) \\
\hline
\end{tabular}
\end{center}
\end{table}

\begin{table}
\caption{Case study on COPD: Graph measures results, including number of edges, global clustering, betweenness centrality and count of hub nodes, for each subgroup. Hub nodes are defined as nodes with a degree $\geq 4$, or at least four connections.  Specific hub nodes and extended degree results can be found in the Supplementary Material.}
\label{t:two2}
\begin{center}
\vskip 2mm
{\bf Fc$\gamma$R pathway}\\
\begin{tabular}{c c c c c c c}
\multicolumn{3}{c}{\bf Metabolites}&&\multicolumn{3}{c}{\bf Genes}\\
\hline\hline	
Group 1&Group 2& Group 3 & & Group 1 & Group 2 & Group 3\\ \hline 
59&57&58 & Number of Edges & 405&444&332\\
0.1665&0.2430&0.1683 & Global Clustering &0.4268&0.4495&0.4442\\
0.2122&0.2783&0.3348 & Betweenness Centrality & 0.0771&0.0483&0.0995\\
12 & 5 & 10 & Count of Hub Nodes & 50 & 53 & 46\\
\hline\hline
\end{tabular}
\vskip 5mm
{\bf Reg Auto pathway }\\
\begin{tabular}{c c c c c c c}
\multicolumn{3}{c}{\bf Metabolites}&&\multicolumn{3}{c}{\bf Genes}\\
\hline\hline	
Group 1&Group 2& Group 3 & & Group 1 & Group 2 & Group 3\\ \hline 
49&51&54 & Number of Edges &71&76&49 \\
0.0881&0.2143&0.1003 & Global Clustering & 0.4649&0.5175&0.4123\\
0.1524&0.21117&0.1862 & Betweenness centrality&0.1800&0.1205&0.1435\\
9 & 8 & 6  & Count of hub nodes & 14 & 15 & 8\\
\hline\hline
\end{tabular}
\end{center}
\end{table}

\begin{table}
\begin{center}
\caption{Simulation study: In setting one, one group on one of the two platforms is dissimilar from the others. In setting two, both platforms have dissimilar groups. Network accuracy metrics are reported as Mean (Standard Error) over 25 replicates for $p = 80$ scenarios and 50 replicates for $p = 40$ scenarios.}
\label{tab:sim}
\vskip 5mm
{\bf Setting one,}  $p=40$
\begin{tabular}{ l c  c  c  c }
\hline
		{\bf{Method}} &{\bf{ TPR }}& {\bf{FPR  }}&{\bf{ MCC }} &{\bf{ AUC }}\\ \hline 			
		Fused Lasso & 0.743 (0.0031) &	0.028 (0.0004) &	0.639 (0.0028) & 0.936 (0.0020) \\
                 Group Lasso & 0.785 (0.0031) &	0.060 (0.0005) &	0.536 (0.0023) & 0.912 (0.0022) \\
		Hub Group Lasso & 0.123	(0.0049) & 0.005 (0.0004) &	0.263 (0.0062) & 0.899 (0.0022) \\	
		Multi-Platform Bayes & 0.611 (0.0063) &	0.022 (0.0005) & 0.579 (0.0055) & 0.895 (0.0036)\\
		\hline
		\hline
\end{tabular}
\vskip 7mm
{\bf Setting two,} $p=40$\\
\begin{tabular}{ l  c  c  c  c }		
\hline
		{\bf{Method}} &{\bf{ TPR }}& {\bf{FPR }}&{\bf{ MCC }} &{\bf{ AUC }}\\ \hline			
		Fused Lasso & 0.907	(0.0016) & 0.157	(0.0006) &	0.439	(0.0012) & 0.963 (0.0003)\\
                Group Lasso & 0.930	(0.0015) & 0.167	(0.0005) & 0.436 (0.0010)  & 0.954 (0.0004)\\
		Hub Group Lasso& 1.000	(0.0000) & 0.467	(0.0048) & 0.232 (0.0021) &0.945 (0.0004)\\
		Multi-Platform Bayes & 1.000	(0.0001) & 0.028	(0.0004) & 0.794 (0.0020) & 1.000 (0.0001)\\
		\hline\hline
\end{tabular}
\vskip 7mm
{\bf Setting one,} $p=80$\\
\begin{tabular}{ l  c  c  c  c }		
\hline		
		{\bf{Method}} &{\bf{ TPR  }}& {\bf{FPR }}&{\bf{ MCC }} &{\bf{ AUC }}\\ \hline			
		Fused Lasso & 0.657	(0.0037) & 0.035	(0.0005) & 0.546 (0.0029)  & 0.919 (0.0023)\\
                 Group Lasso & 0.777	(0.0031) & 0.069	(0.0005) & 0.506	(0.0022)  & 0.916 (0.0023)\\
		Hub Group Lasso& 0.263	(0.0064) & 0.005	(0.0003) & 0.427  (0.0050) & 0.905 (0.0023)\\
		Multi-Platform Bayes & 0.636	(0.0063) & 0.023	(0.0005) & 0.597 (0.0053) & 0.941 (0.0037)\\ 
		\hline \hline 
\end{tabular}
\vskip 7mm
{\bf Setting two,} $p=80$\\
\begin{tabular}{ l  c  c  c  c }		
\hline		
		{\bf{Method}} &{\bf{ TPR  }}& {\bf{FPR }}&{\bf{ MCC }} &{\bf{ AUC }}\\ \hline			
		Fused Lasso & 0.735	(0.0017) & 0.080	(0.0004) & 0.451 (0.0015) &  0.957 (0.0009)\\
                 Group Lasso & 0.998	(0.0002) & 0.270	(0.0006) & 0.343	(0.0005)  & 0.938 (0.0229)\\
		Hub Group Lasso& 1.000	(0.0000) & 0.464	(0.0044) & 0.233 (0.0019) & 0.945 (0.0004)\\
		Multi-Platform Bayes & 1.000	(0.0001)& 0.026	(0.0004) & 0.808 (0.0021) & 1.000 (0.0001)\\ 
		\hline \hline 
\end{tabular}
\end{center}
\end{table}
  
\end{document}